# Hubble Space Telescope Pre-Perihelion ACS/WFC Imaging Polarimetry of Comet ISON (C/2012 S1) at 3.81 AU

Dean C. Hines,[1,2] Gorden Videen,[2,3] Evgenij Zubko,[4,5] Karri Muinonen,[4,6] Yuriy Shkuratov,[5,7] Vadim G. Kaydash,[5] Matthew M. Knight,[8,9] Michael L. Sitko,[2,10] Carey M. Lisse,[9] Max Mutchler,[1] Derek Hammer,[1] Padmavati A. Yanamandra-Fisher[2]

[1] Space Telescope Science Institute, Baltimore, MD 21218 USA

[2] Space Science Institute, 4750 Walnut Street, Suite 205, Boulder, CO 80301, USA

[3] U.S. Army Research Laboratory, 2800 Powder Mill Road, Adelphi, MD 20783, USA

[4] Department of Physics, PO. Box 64, FI-00014 University of Helsinki, Finland

[5] Astronomical Institute of V.N. Karazin University, Kharkov, 61058, Ukraine

[6] Finnish Geodetic Institute, PO. Box 15, FI-02431 Masala, Finland

[7] Radioastronomical Institute of NASU, Kharkov, 61002, Ukraine

[8] Lowell Observatory, Flagstaff, AZ 86001 USA

[9] Johns Hopkins University, Applied Physics Laboratory, Laurel, MD 20723 USA

[10] University of Cincinnati, Department of Physics, Cincinnati, OH 45221 USA

**Abstract**

We present polarization images of Comet ISON (C/2012 S1) taken with the *Hubble Space Telescope (HST)* on UTC 2013 May 8 ($r_h$ = 3.81 AU, $\Delta$ = 4.34 AU), when the phase angle was $\alpha \approx 12.16°$. This phase angle is approximately centered in the negative polarization branch for cometary dust. The region beyond 1000 km (~0.32 arcseconds ≈ 6 pixels) from the nucleus shows a negative polarization amplitude of p% ~-1.6%. Within 1000 km of the nucleus, the polarization position angle rotates to be approximately perpendicular to the scattering plane, with an amplitude p%~+2.5%. Such positive polarization has been observed previously as a characteristic feature of cometary jets, and we show that Comet ISON does indeed harbor a jet-like feature. These *HST* observations of Comet ISON represent the first visible light, imaging polarimetry with sub-arcsecond spatial resolution of a Nearly Isotropic Comet (NIC) beyond 3.8 AU from the Sun at a small phase angle. The observations provide an early glimpse of the properties of the cometary dust preserved in this Oort-cloud comet.

# 1. Introduction

Since its discovery in September 2012, Comet ISON (C/2012 S1), hereafter Comet ISON, has been recognized as a multi-faceted research opportunity. Its near-parabolic trajectory implies that it is a Nearly Isotropic Comet (NIC), approaching the inner solar system from the Oort cloud for the first time since its formation 4.5 Gya. Thus, it contains a fossil record of the dusty volatile material that existed in the Solar System within a few million years after the collapse of the nebula from which it formed. As Comet ISON approaches the Sun, some of this material will be expelled from the surface, enabling detailed compositional studies. Comet ISON is a sun-grazing comet, and is expected to show significant activity if it survives this encounter.

Studying cometary dust is challenging. Currently, *in-situ* capture of dust grains (e.g., Stardust: Brownlee et al. 2004, 2006) is limited to a single comet, and the fragile grains disaggregated upon entry into the aerogel, so that the original structure was destroyed (e.g., Zolensky et al. 2006). Cometary dust grains may be captured by the Earth atmosphere-terrestrial system. Fragile porous grains collected in the upper stratosphere often survive atmospheric entry, but still undergo some heating during deceleration and possibly some damage upon capture (Messenger 2003). The specific comet of their origin is difficult to determine due to background contamination by unrelated interplanetary dust particles. Hence, there is still a paucity of reliable structural information on cometary grains whose point of origin can be identified. We need remote observations of specific objects, and every comet that can be studied in this manner improves our understanding of these grains.

In addition to direct imaging and spectroscopy (e.g., Hanner et al. 1994; Lisse et al. 1998, 2004, 2006; Wooden et al. 1999; Woodward et al. 2007; Kelley & Wooden 2009), imaging polarimetry provides a powerful tool to investigate the detailed structural nature of cometary dust grains (see, e.g., reviews by Kolokolova et al. 2004; Levasseur-Regourd et al. 2008). The polarizing properties of grains depend on the angle through which light is scattered. Astronomers often use the Sun-Target-Observer (STO) angle (i.e., phase angle $\alpha$) of the observations, which is related to the physical scattering angle via $\alpha=180°$–scattering angle. Maximum polarization occurs at $\alpha \sim 90\text{-}110°$, and the dominant electric-vector plane of the scattered light, hereafter the "plane of polarization," is perpendicular to the STO scattering plane. However, for $\alpha <\sim 20°$, the plane of polarization can lie in the STO plane, a phenomenon referred to as negative polarization (e.g., Kiselev & Chernova 1978; Johnson et al. 1980; Levasseur-Regourd et al. 1996; Yanamandra-Fisher & Hanner 1999; Videen et al. 2004; Muinonen et al. 2007; Shkuratov et al. 2011; Mishchenko et al. 2010; Shkuratov et al. 2011).

Ground-based polarization images of comets obtained at various phase angles $\alpha$, including within the negative polarization branch ($\alpha$ below about 23°), usually show significant changes through the coma, indicating an inhomogeneous distribution of grains (Renard et al. 1996; Hadamcik & Levasseur-Regourd 2003a,b; Levasseur-Regourd & Hadamcik 2003). While much of the coma is often positively (or slightly negatively) polarizing, the innermost region, called the circumnucleus halo, can have a large negative polarization (~-6%) at small phase angles ($\alpha \sim 10\text{-}15°$). This implies that particles in the circumnucleus halo must have properties (i.e., composition, shape/size, or orientation) different from other particles in the coma. Images from some comets, i.e. 22P/Kopff, 81P/Wild 2, C/1990 K1 Levy, C/1995 O1 Hale-Bopp, show this circumnucleus polarimetric halo extending extending some ~500-5,000 km (Hadamcik & Levasseur-Regourd 2003b). In addition to the large negative polarization seen in the circumnucleus halo region, cometary jets appear to have a positive polarization signal, indicating yet another population of dust particles.

Characterizing the coma with low-spatial-resolution observations is problematic, because these features become beam-diluted and washed-out, and the net measured polarization may not represent any specific portion of the cometary coma. The measured polarization may also be affected by light scattered from particles in the cometary tail, which at small $\alpha$ could have a significant component projected along line-of-sight through the coma.

Comet ISON presents a tremendous polarimetric-imaging opportunity to examine comet heterogeneity. Previous observations of both short-period and long-period comets (Hadamcik & Levasseur-Regourd 2003a,b) have consistently shown a polarimetric circumnucleus halo reaching ~-6%. Based on numerical light-scattering simulations of agglomerated particles, Zubko et al. (2012) suggested that such high negative polarization is consistent with depletion of highly absorbing carbonaceous materials, which could result from, e.g., processing by radiation for extended periods. This exposure can cause sputtering and photolysis reactions of carbonaceous materials, ablating them and producing less absorbing chemical species. The circumnucleus halo particles also could be associated with a so-called *crust*, an outermost layer of refractory materials remaining on the surface of the cometary nucleus after sublimation-loss of volatiles (e.g., Whipple 1950). The crust does not necessarily contain just primordial materials; it may have been modified by external effects (e.g., heating and space weathering). Because Comet ISON is expected to be a fresh comet newly arrived from the Oort cloud, its crust may not be developed in the same way as a short-period comet that has experienced previous insolation, and we might expect that it could reveal unusual polarimetric properties within the circumnucleus halo.

Here we present polarimetric images of Comet ISON captured with the *Hubble Space Telescope (HST)* on UTC 2013-May-8, when the phase angle was $\alpha \approx 12.16º$; i.e., where the negative polarization amplitude is expected to be largest. This small phase angle occurred when the comet was approximately 3.81 AU from the Sun (4.34 AU from the Earth), beyond the pure water-ice sublimation distance. Our results provide the first polarimetric observations of such a distant NIC at a small phase angle with sub-arcsecond spatial resolution.

## 2. Observations and Data Reductions

Two orbits of Director's Discretionary Time (DD/GO 13199: P.I. D. Hines) were used to observe Comet ISON with the Wide Field Channel (WFC) of the Advanced Camera for Surveys (ACS: Ford et al. 1998) aboard *HST*. The orbital elements for Comet ISON were from the JPL HORIZONS System; solution JPL#20, 2013-Mar-24 20:50:46. The three visible-band polarizers were used in conjunction with the F606W (broad V) filter to obtain six images of the comet; two images offset by 3.011″ were obtained through each polarizer to mitigate star trails, cosmic rays, bad pixels and other residual image artifacts. The total on-source integration time per polarizer/filter (hereafter POL*V) image was 1498 seconds. Table 1 presents a log of the observations.

## 2.1. Basic Reductions

Raw ACS/WFC images were processed into total count images using the standard *calacs* pipeline at the Space Telescope Science Institute. Sky background was estimated using the median value from an iteratively sigma-clipped region away from the comet and vignetted areas. These sky values were then inserted into the MDRIZSKY header keyword, and subtracted by *AstroDrizzle* (Gonzaga, et al., 2012) during generation of images that are corrected for the significant field distortions imposed by the off-axis configuration of the ACS/WFC relative to the telescope bore-sight. This step is crucial; if not corrected properly, the distortions can lead to spurious polarization signatures. We checked the efficacy of the distortion corrections by reducing archival images of the Egg Nebula (CRL 2688) using exactly the same procedure as for Comet ISON. The Egg Nebula is particularly informative because it exhibits highly polarized, nearly perfect centrosymmetric emission shells, and is characterized well from *HST*/NICMOS polarimetry (Sahai et al. 1998; Hines et al. 2000; Weintraub et al. 2000). The *AstroDrizzled* ACS/WFC polarization images of the Egg Nebula show precise centrosymmetric structure, and the perpendicular (pseudo-)vectors converge to a point as expected (Hines et al. in prep). We conclude that *AstroDrizzle* is removing the field distortions of ACS/WFC POL*V/F606W correctly.

The comet nucleus centroid was used to align the images. A single count-rate image was formed from the two images per POL*V, using *AstroDrizzle* with *EXP* weighting to remove (by rejecting the brightest pixels) star trails, residual cosmic rays and bad pixels. No correction for charge transfer efficiency (CTE) was made, since the exposures for each image provide enough background signal to mitigate CTE loss in regions of interest; we expect no effect on the polarization, and indeed no effects are seen in the Egg Nebula polarimetry.

## 2.2. Formation of Stokes Images and Measurement of the Polarization

To construct Stokes parameter images, each POL*V image was shifted linearly into a common reference frame. The comet had high proper motion, so alignment using background (unpolarized) stars was impossible. Instead, centroids from two-dimensional Gaussian fits to the 20 pixels surrounding the comet nucleus were used to align the POL*V images. We tested the alignment efficacy by evaluating differenced image pairs with slight offsets from the centroid position. Shifts ≥ 0.25 pixels caused clearly unphysical artifacts; our alignment is correct to within this 0.25 pixel tolerance. Our extensive experience shows that polarimetric measurement of features finer than 4x pixel-scale can be unreliable, so the images were smoothed with a 5x5 pixel boxcar prior to forming the Stokes images. Finally, we performed aperture polarimetry measurements on the combined Stokes images and on Stokes images constructed from the two images per POL*V independently, with consistent results.

Stokes parameters are computed from the three POL*V images via the transformation for an *ideal*, three-polarizer system:

$$I = \frac{2}{3} (I_0 + I_{60} + I_{120}), \quad (1)$$

$$Q = \frac{2}{3} (2I_0 - I_{60} - I_{120}), \quad (2)$$

$$U = \frac{2}{\sqrt{3}} (I_{60} - I_{120}). \quad (3)$$

The percentage of linear polarization is given by $p = 100\%(Q^2 + U^2)^{1/2}/I$, and the polarization position angle in the instrument frame is given by $\theta_{instr} = 0.5\tan^{-1}(Q/U)$.[1] This angle placed in the celestial frame is $\theta_p = \theta_{instr} - 38.15° + PAV3$, where PAV3 is the position angle of the *HST* V3 axis during observations (Biretta & Kozhurina-Platais 2004; Cracraft & Sparks 2007).

---

[1] Assumes a 360° arctangent function.

Importantly, the ACS visible polarizers are *not ideal*, and there is a flat mirror in the ACS/WFC optical train (Biretta et al. 2004). Even so, Cracraft & Sparks (2007) found that multiplicatively scaling the POL*V images removes instrumental signatures for objects with intrinsic polarizations of p%≥5-10% (Sparks et al. 2008). We expected the Comet ISON polarization to be "low," p%≤6%. Therefore, we analyzed ACS polarimetry observations of an unpolarized star (GD319: Turnshek et al. 1990: Schmidt et al. 1992) and the polarized standard Vela I (No. 81: Whittet et al. 1992) obtained in program CAL 10055 (P.I. J. Biretta). We find that the coefficients listed in Cracraft & Sparks (2007) reproduce: 1) the polarization of Vela I to within 0.3% and $\theta_p$ to within ~3°; and 2) a null result for GD319 to within p%≈0.3%. To calculate the *corrected* Stokes parameters, we apply the following scaling $I(cor)_{POL*V} = C_{POL*V} * I(obs)_{POLV}$, where $C_{POL0V} = 1.2960$, $C_{POL60V} = 1.3238$, $C_{POL120V} = 1.2781$, and then use equations 1-3.

## 3. Results

Figure 1 shows the three *AstroDrizzled* POL*V images, plus a stack of all of the images, but without cosmic ray or star-trail removal; this is useful for spotting regions contaminated by residual cosmic rays and star trails, which should only be used with extreme caution in the polarimetric analysis.

Figure 2 shows the total intensity image of Comet ISON with polarization (pseudo-)vectors overlaid in regions where the S/N in p% is ≥5.[2] The coma polarization is approximately constant at p% ≈1.6% with the position angle on the sky $\theta_p$ ≈ 92°, but the position angle rotates by ~64°, to $\theta_p$ ≈ 156° at the center 5x5 pixel bin. At the time of the observations, the Sun was at position angle PA=269.74°, which places the scattering plane at 179.74° (which is the positive-polarization plane).

To better determine the polarization profile, we measured the Stokes parameters (in the scattering plane) in successive annuli from the peak in the total intensity image. Figure 3 shows: (a) *q* versus *u*; and (b) *q* & *u* as a function of radius from the peak in the total intensity image. Uncertainties were estimated via total counts (electrons) per annulus per image $\sigma_q=\sigma_u=\sigma_p\sim\sqrt{2/Total\ Counts}$ ~0.3%, where $p/\sigma_p$>5, added in quadrature with the instrumental calibration uncertainty ($\sigma_p\%_{inst}$=0.3%). Uncertainties in the position angle are given by $\sigma_\theta$~28.65°($\sigma_p$(tot)/p)~8°.

Light scattered from the coma is clearly polarized, showing negative polarization expected for comets at similar phase angles (Dollfus et al. 1988; Hadamcik & Levasseur-Regourd 2003a,b; Levasseur-Regourd 2003; Kelley et al. 2004). However, within ~0.318″ (=1000 km) of the coma center[3], $\theta_p$ rotates and the polarization becomes increasingly positive, approaching p%~+2.5%. Image misalignment might cause this, especially for the very steep intensity gradients near the bright nucleus. *However, the rotation was found even when computing the Stokes parameters from multi-aperture photometry of all six polarizer images.*

---

[2] The degree of polarization was not bias-corrected (e.g., Wardle & Kronberg 1974; Simmons & Stewart 1985), because we restrict ourselves to p/$\sigma_p$ ≥ 5.
[3] In the 5x5 pixel, boxcar-smoothed data.

## 4. Discussion

A primary objective for observing Comet ISON was to resolve the circumnucleus halo of an Oort-cloud comet beyond the pure water-ice line, for comparison with short-period comets. Some observed polarization properties of Comet ISON are typical of other comets observed at similar phase angles, even for objects observed much closer to the Sun, including a change in $\theta_p$ suggesting that scattering particles within a few hundred kilometers of the nucleus have different properties (either compositionally, structurally, or in orientation) compared with more distant material. However, the lack of a circumnucleus halo region with high negative-polarization (~-6%) in Comet ISON seems not to be what was expected.

Dilution by unpolarized gas emission could cause lower-than-expected negative polarization. While CO and $CO_2$ sublimate at this solar distance, these molecules do not emit significantly within the F606W band-pass. Also, there are no contemporaneous reports of emission from gas species, such as $C_2$, $NH_2$, or [OI], that would fall within the F606W band-pass. By far, $C_2$ would be the strongest contributor, so following Sen et al. (1989) we estimate the $C_2$ polarization at α=12.16° to be p%~+0.18%. Even if the $C_2$ equivalent-width is a significant fraction of the F606W band-pass (Δλ≈2000Å), the intrinsic negative polarization (from dust-scattered light) would only increase to ~-1.8%. We conclude that the lower-than-expected negative-polarization is a manifestation of the scattering particle properties, and not related to dilution by (positive) polarization from resonant-scattered molecular emission.

The other interesting polarimetric feature of Comet ISON is the measurement of a positive polarization (p%~+2.5%) component within 1000km. While irregularly shaped particles can produce negative polarizations approaching p%=-10% at these phase angles, they do not produce positive polarization (Muiononen et al. 2002; Zubko et al. 2009: Muinonen et al. 2012). Such positive polarizations can arise by scattering from icy grains and regularly shaped particles (spheres or spheroids). Optically thin, Rayleigh scattering, for example, reaches $p=100\%*\sin^2\alpha/(1+\cos^2\alpha)=+2.3\%$ at $\alpha=12.16°$. However, in that case the position angle should be perpendicular to the scattering plane, yet we measure $\Delta\theta\sim64°$.

This discrepancy could indicate the presence of a polarized component with intermediate position angle. Some constraints can be placed by vectorally subtracting the negative-polarization component from the measurements; i.e., considering the negative-polarzation component as a "sky-background." A polarization component ~4% oriented ~78° relative to the negative-polarization component would produce a net signal ~2.5% oriented at ~64° relative to the negative-polarization "background." Such a "u" component could suggest scattering from an optically thick structure (see, e.g., Zubko & Laor 2000). A combination of optically thin (pure positive-polarization) and optically thick scattering components that are unresolved spatially is also a possibility. Finally, this may also indicate some residual systematic error at the very center that we cannot eliminate completely (but see §3). Regardless, there is apparently a scattering component within 1000km with polarization properties different than the negative-polarization component.

Jet features in comets have exhibited positive polarization (Hadamcik & Levasseur-Regourd 2003a,b; Levasseur-Regourd 2003; Hadamcik et al. 2013). *HST* Wide Field Camera 3 (WFC3) images of Comet ISON obtained on UT 2013-APR-10 show the coma generally follows the expected 1/ρ brightness distribution (ρ is the radial distance from the nucleus: Gehrz & Ney 1992), but a jet-like asymmetry is also seen (Li et al. 2013). Our observations reveal similar morphology compared with the WFC3 images (Figure 4). Scattering off particles in this jet-like feature might contribute to the positive polarization observed within 1000km of the nucleus.

Previous numerical light-scattering simulations suggest that negative-polarization in circumnucleus halos can result from depletion of absorbing particles (Zubko et al. 2009; Zubko et al. 2012), which could be due to processing of carbonaceous material in this portion of the coma. Furthermore the positive-polarization signal very near the nucleus suggests abundant icy grains, a phenomenon seen in the recent close flyby of 103P/Hartley 2 (A'Hearn et al. 2011). These icy grains should be ephemeral, as they evaporate in sunlight, hence their localization near the nucleus, but could also be partially responsible for the lower negative-polarization in the circumnuclear halo. This may explain the "bluer" color of the circumnuclear region compared with the rest of the coma as observed in the WFC3 images (Li et al. 2013).

## 5. Conclusions

We present polarimetric images of Comet ISON captured with the *HST* ACS/WFC at a phase angle α=12.16°, near the maximum of the negative polarization branch. The average (negative) polarization over the coma is p% ≈-1.6%. Unlike some other short-period comets, a strong negative-polarization circumnucleus halo is not observed. Instead, a positive-polarization component appears to exist within a few hundred kilometers of the nucleus, with a measured polarization of p%~+2.5%, possibly associated with the observed, extended jet-like feature. A strong negative-polarization circumnucleus halo could indicate a depletion of absorbing particles in this region (Zubko et al. 2012). Therefore, the lack of this halo suggests the presence of absorbing particles, or icy grains and particles too small to produce a negative-polarization branch.

Our observations were obtained when the comet was beyond the water-ice sublimation distance. As the comet continues its orbit, we might expect the unique (polarimetric) features to disappear. It will be interesting to compare our observations with additional polarimetric images obtained at later epochs and post-perihelion, assuming the comet survives this encounter.

## 5. Acknowledgements

We wish to thank A. C. Levasseur-Regourd for a careful reading of the manuscript, and for suggestions that have enhanced its clarity. We also thank L. Kolokolova for useful discussions. We also acknowledge the anonymous referee for comments that have improved the manuscript significantly. Support for program number GO 13199 was provided by NASA through a grant from the Space Telescope Science Institute, which is operated by the Association of Universities for Research in Astronomy, Inc., under NASA contract NAS5-26555. Additional funding by the SAEMPL ERC Advanced Grant No. 320773 (Scattering and Absorption of Electromagnetic Waves in Particulate Media) and the Academy of Finland contract No. 257966.

**Table 1:** Observation Log

| Comet Right Ascension (J2000.0) | Comet Declination (J2000.0) | UTC Start Time | Comet Phase Angle[a] (degrees) | On-Source Exposure Time (sec) | Polarizer/Filter | MAST[b] Archive Data Set |
|---|---|---|---|---|---|---|
| 06 44 31.133 | +29 11 45.83 | 2013-05-07 19:47:16 | 12.170 | 1498.000 | POL0V/F606W | JC7F01010 |
| 06 44 33.023 | +29 11 36.53 | 2013-05-07 21:23:21 | 12.168 | 1498.000 | POL60V/F606W | JC7F01020 |
| 06 44 35.957 | +29 11 22.11 | 2013-05-07 23:53:48 | 12.157 | 1498.000 | POL120V//F606W | JC7F01030 |

[a] at beginning of observation (JPL Horizons)

[b] Mikulski Archive for Space Telescopes (MAST) at the Space Telescope Science Institute (STScI).

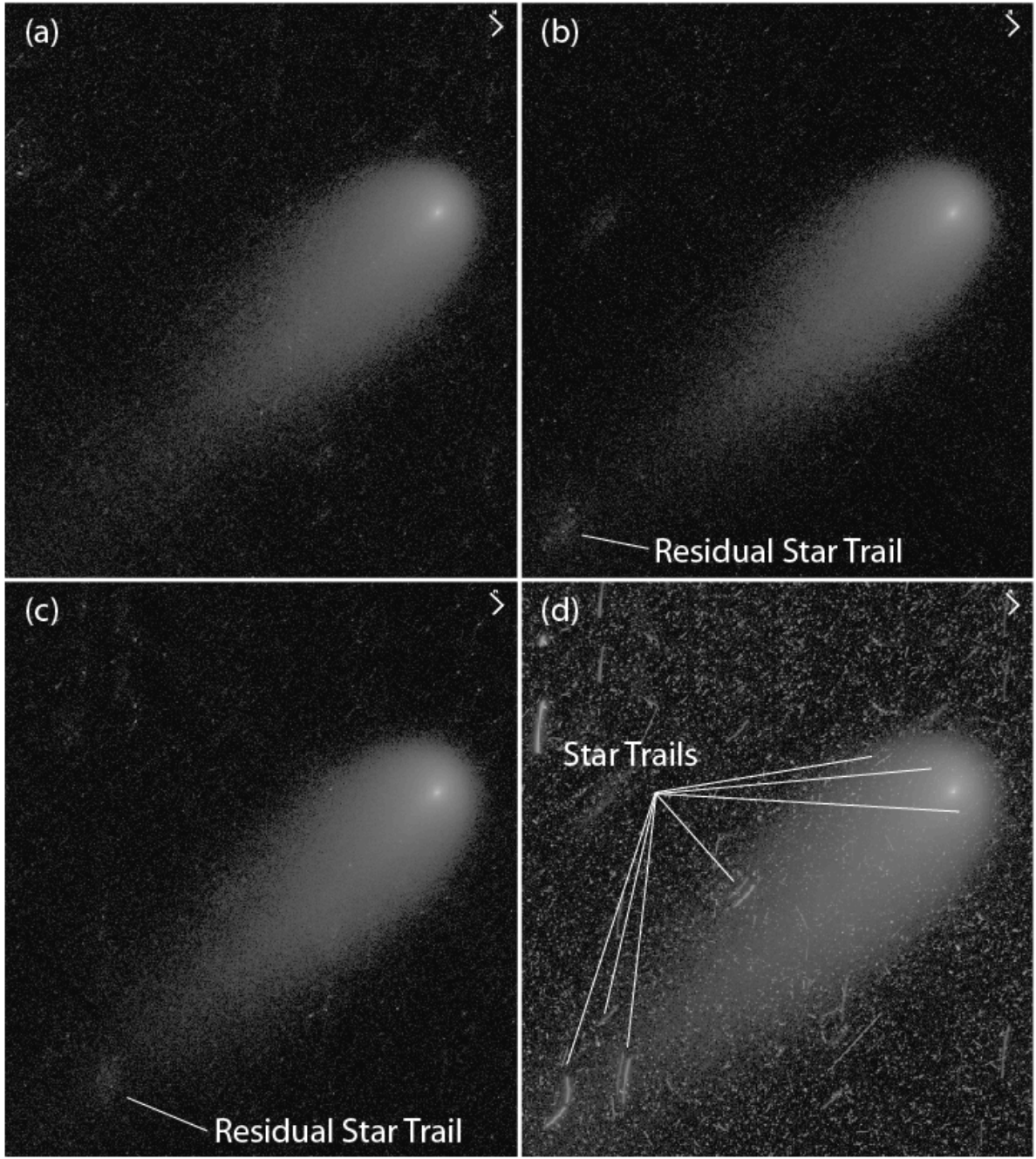

**Figure 1:** ACS/WFC polarizer/filter images of Comet ISON in detector coordinates, with logarithmic stretch. (a) POL0V/F606W, (b) POL60V/F606W, (c) POL120V/F606W, (d) stack of all three images, but without cosmic ray or star-trail rejection. The stack shows the locations of trailed stars that may affect the final polarimetric analysis. In particular, the two blobs behind the comet tail in (b) and (c) are residuals.

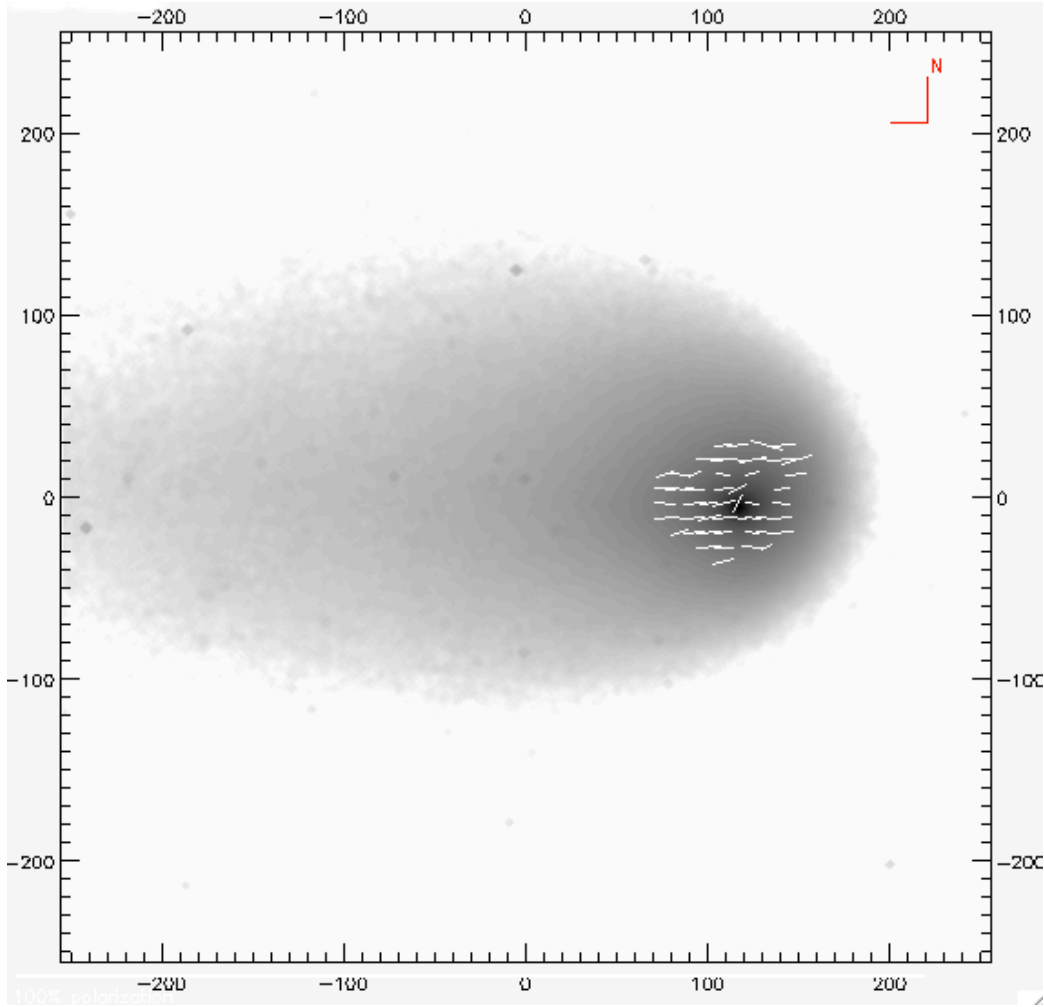

**Figure 2:** Boxcar-smoothed (5x5pixel) imaging polarimetry of Comet ISON showing magnitude (inverted grey scale) and direction of the polarization vectors near the center; x & y axes in detector pixels (0.05″/pix). A 100% polarized (pseudo-)vector would extend 500 pixels in the diagram. The image is displayed in celestial coordinates, North Up and East Left; the Sun is due West (PA=269.74°).

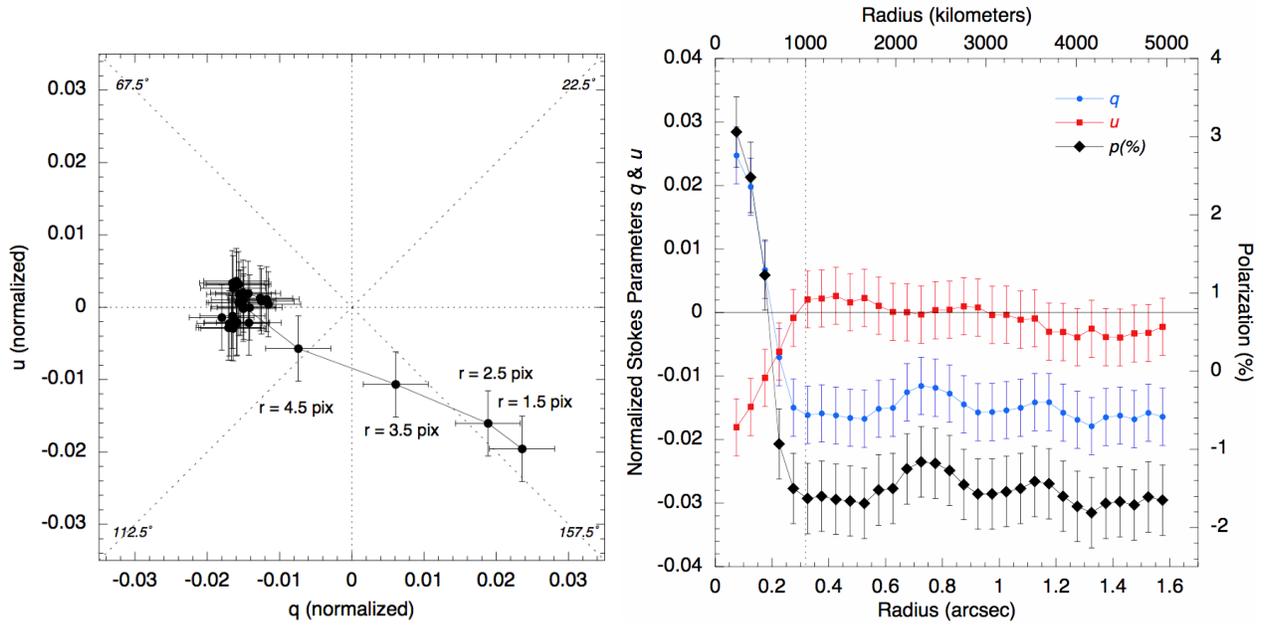

**Figure 3:** Imaging polarimetry of Comet ISON as a function of distance from the central peak brightness showing: (left) normalized *q vs u*. (+*q* is perpendicular to the scattering plane); (right) normalized *q&u* and percentage polarization.

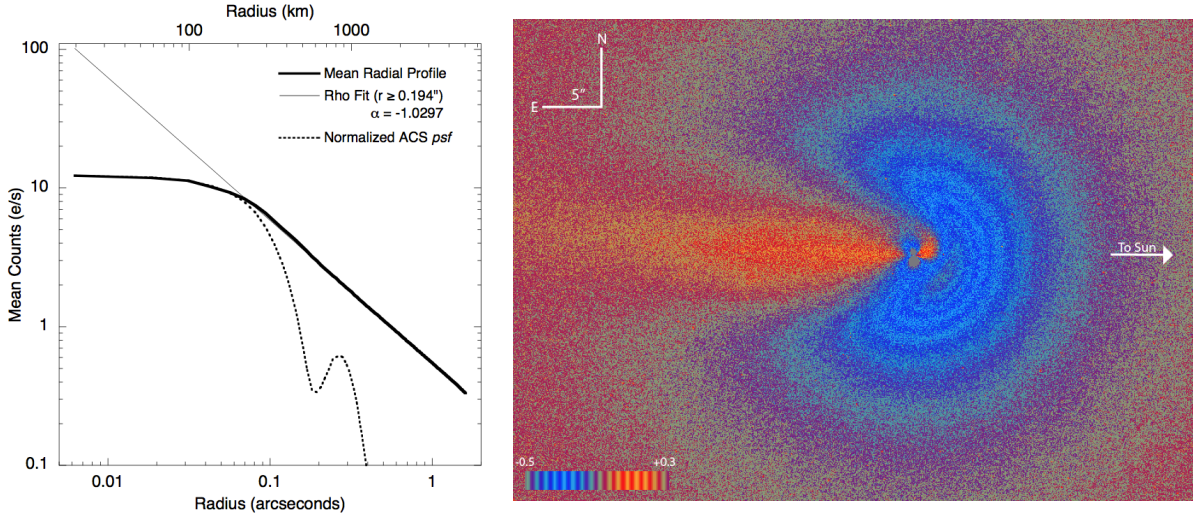

**Figure 4:** (left) Radial profile of the total intensity of Comet ISON compared with that of a model of the point-source-function for the ACS/WFC F606W filter using a G2V stellar spectrum. The radial profile beyond 200km is well fit by a $1/\rho$ profile. (right) The total intensity image (in e s$^{-1}$) after subtraction of the $1/\rho$ model, revealing an asymmetric jet-like feature in the sunward direction.